\documentclass[aps,pra,twocolumn,superscriptaddress]{revtex4-2}

\usepackage{graphicx}
\usepackage{subfigure}
\usepackage{amsmath}
\usepackage{mathrsfs}
\usepackage{amsthm}
\usepackage{amssymb}
\usepackage{hyperref}
\usepackage{xcolor}
\usepackage[OT4]{fontenc}
\usepackage[utf8]{inputenc}

\usepackage{ulem}
\usepackage{dcolumn}
\usepackage{bm}

\hypersetup{colorlinks=true, linkcolor=black, citecolor=black, urlcolor=blue}
\allowdisplaybreaks

\begin{document} 

\newcommand{\ketbra}[2]{|#1\rangle\!\langle#2|}
\newcommand{\bra}[1]{\langle #1|}
\newcommand{\ket}[1]{| #1 \rangle}
\newcommand{\braket}[2]{\langle #1 | #2 \rangle}
\newcommand{\braketbig}[2]{\big\langle #1 \big| #2 \big\rangle}
\newcommand{\braketbigg}[2]{\bigg\langle #1 \bigg| #2 \bigg\rangle}
\newcommand{\tr}[1]{{\rm Tr}\left[ #1 \right]}
\newcommand{\av}[1]{\langle{#1}\rangle}
\newcommand{\avbig}[1]{\langle{#1}\big\rangle}
\newcommand{\avbigg}[1]{\bigg\langle{#1}\bigg\rangle}
\newcommand{\bk}{\mathbf{k}}
\newcommand{\bp}{\mathbf{p}}
\newcommand{\re}{\mathrm{Re}}
\newcommand{\im}{\mathrm{Im}}

\newcommand{\tw}[1]{{\color{blue} [TW: #1]}}
\newcommand{\kj}[1]{{\color{magenta} [#1]}}
\newcommand{\twm}[1]{{\color{blue} #1}}
\newcommand{\twr}[2]{{\color{blue} \sout{#1} #2}}

\newcommand{\x}{{\bf r}}
\newcommand{\K}{{\bf k}}
\newcommand{\dk}{\Delta {\bf k}}
\newcommand{\DK}{\Delta {\bf K}}
\newcommand{\KK}{{\bf K}}
\newcommand{\X}{{\bf R}}
\newcommand{\B}[1]{\mathbf{#1}} 
\newcommand{\f}[1]{\textrm{#1}} 
\newcommand{\half}{{\frac{1}{2}}}
\newcommand{\vv}{{\bf v}}
\newcommand{\p}{{\bf p}}
\newcommand{\dx}{\Delta {\bf r}}
\newcommand{\q}{{\bf q}}

\author{Maciej Pylak}
\affiliation{Institute of Physics, Polish Academy of Sciences, Aleja Lotnik\'ow 32/46, PL-02-668 Warsaw, Poland}
\affiliation{National Centre for Nuclear Research, ul. Pasteura 7, PL-02-093 Warsaw, Poland}

\author{Filip Gampel}
\affiliation{Institute of Physics, Polish Academy of Sciences, Aleja Lotnik\'ow 32/46, PL-02-668 Warsaw, Poland}
\email{gampel@ifpan.edu.pl}

\author{Marcin P{\l}odzie\'n}
\affiliation{Institute of Physics, Polish Academy of Sciences, Aleja Lotnik\'ow 32/46, PL-02-668 Warsaw, Poland}
\affiliation{ICFO, The Barcelona Institute of Science and Technology, Av. Carl Friedrich Gauss 3, 08860 Castelldefels (Barcelona), Spain}

\author{Mariusz Gajda}
\affiliation{Institute of Physics, Polish Academy of Sciences, Aleja Lotnik\'ow 32/46, PL-02-668 Warsaw, Poland}
\email{gajda@ifpan.edu.pl}

\title{Manifestation of relative phase in dynamics of two interacting  Bose-Bose droplets}

  \begin{abstract}
We study coherent dynamics of two interacting Bose-Bose droplets by means of the extended Gross-Pitaevskii equation. The relative motion of the droplets couples to the phases of their components. The dynamics can be understood in terms of the evolution of zero-energy modes recovering symmetries spontaneously broken by the mean-field solution. These are translational symmetry and two U(1) symmetries, associated with the phases of the droplets’ two components. A phase-dependent interaction potential and double Josephson-junction equations are introduced to explain the observed variety of different scenarios of collision. We show that the evolution of the droplets is a macroscopic manifestation of the hidden dynamics of their phases. The occurrence of nondissipative drag between the two supercurrents (Andreev-Bashkin effect) is mentioned.

\end{abstract}

\maketitle

\section{Introduction}

Quantum droplets are self-bound objects formed by ultracold atoms. Despite having densities about eight orders of magnitude smaller than air they behave like liquids. Droplets were first observed in dipolar gases in systems of dysprosium \cite{Pfau16,Pfau16a, Pfau16b}  or erbium atoms \cite{Ferlaino16}. Their binding mechanism occurred to be the same as predicted by Petrov \cite{Petrov15} for two-component Bose-Bose mixtures. Self-bound systems of ultracold atoms are formed when the mean-field energy of the gas almost vanishes and quantum fluctuations become important. The Lee-Huang-Yang \cite{LHY,Fischer,Pelster11,Pelster12}  contribution to the energy of the system constitutes an essential ingredient of this stabilizing mechanism. Quantum droplets as considered by Petrov \cite{Petrov15} were obtained in a mixture of two hyperfine states of $^{39}$K \cite{Cabrera18,Tarruell18,Semeghini18,G_Ferioli20}. Experiments confirmed the main predictions of the theory: nonspreading density profiles and values of equilibrium densities.

The lifetime of droplets is limited by three-body losses, but in heteronuclear mixtures of $^{87}$Rb and $^{41}$K \cite{Fort19} it might exceed 100~ms \cite{M_Modugno21}, which is much longer than the lifetime of two-component homonuclear mixtures. Therefore experiments with heteronuclear droplets \cite{Fort19} pave a way toward long-lived binary droplets which in turn opens the possibility to study dynamical situations, such as collisions.

In the case of colliding classical droplets, two scenarios are possible, depending on the relative values of kinetic and surface energies: coalescence or splitting \cite{Poo90,Qian97,Pan05}. This was also observed in binary collisions of quantum droplets composed of two hyperfine states of $^{39}$K \cite{Fattori19}. However, the dynamics of these processes may be richer. The superfluid character of the colliding objects introduces additional degrees of freedom: the relative phases of their components. The coupling of two superfluid systems invokes an analogy to the Josephson effect. Coherent Josephson oscillations are at the heart of recently observed supersolid behavior in weakly coupled droplet systems in 1D \cite{Modugno19,Pfau19,Chomaz19}, as well as in two-dimensional geometries \cite{norcia2021,hertkorn21}. A Josephson-junction based approach was used to describe out-of-equilibrium dynamics of a supersolid \cite{Ilzhofer2021}. For a review of recent advances in the
field of quantum droplets see \cite{Pfau18,Petrov18,Ferrier-Barbut19,Malomed19,Bottcher_2021}.

Finally, we want to mention a totally different system, where effects very similar to those studied here were observed. These are coherent oscillations of neutrons during a collision of two different nuclei. This peculiar result was reported recently \cite{Potel,Magierski}. A minute Josephson oscillating current of Cooper paired neutrons was found in analysis of data from collisions of tin and nickel nuclei ${}^{116} {\rm Sn} + {}^{60}{\rm Ni}$ at energies around $150$ MeV, barely sufficient to overcome Coulomb repulsion. As was suggested theoretically \cite{Andreyev17,Bulgac16,Bulgac17,Magierski17}, the phase difference of a \textit{pairing field} of two initially \textit{independent} nuclei is responsible for this effect.

In the present paper we study low-energy collisions of two interacting Bose-Bose droplets taking into account the coherent exchange of atoms between them. Our approach is based on numerical integration of the extended Gross-Pitaevskii (eGP) equations supported by analysis of equations of motion for zero-energy (Goldstone) modes of the system. These two-component Josephson-junction equations allow for deeper understanding of the observed dynamics. We predict a host of different possible scenarios during the droplets’ approach. This includes in particular a coherent transfer of atoms in the form of direct or alternating Josephson currents, which may result in total evaporation of one of the droplets, or their attraction, merging, or repulsion. We observe dynamics akin to a droplet subject to a ponderomotive force, but also the Andreev-Bashkin effect, i.e., entrainment between the two
superfluids.

The paper is organized as follows. In Sec.~\ref{sec:2} we outline the theory of two-component Bose-Bose droplets using eGP equations. Next in Sec.~\ref{sec:3} we discuss the preparation of the initial state. We compare the fully coherent state, where the relative phase between the two droplets is well defined, with two Fock states, where the relative phase is not controlled. In Sec.~\ref{sec:4} we find an expression for the interaction potential between two separate quantum droplets overlapping with their tails. In Sec.~\ref{sec:5} we use a formalism based on the so-called zero-energy modes Hamiltonian to derive Josephson-junction (JJ) equations for the moving two-component junction. The equations allow us to find effective simplified dynamics of a collision of interacting droplets in terms of their relative positions and velocities, coupled to their relative phases and Josephson currents. In Sec.~\ref{sec:6} we present results of time-dependent numerical simulations of collisions for small and large droplets assuming different initial phases of the droplets. These simulations are compared to predictions of the Josephson-junction model which allows for a clear physical picture of the observed dynamics. We conclude with Sec.~\ref{sec:7}.

\section{ Extended Gross-Pitaevskii equations}
\label{sec:2}
A two-component ultracold Bose-Bose droplet may be described quite accurately by a mean-field energy functional. The energy density of a droplet is
\begin{equation}
    \epsilon (n_1,n_2) = \frac{1}{2} \sum_{i,j} n_i g_{ij} n_j+\frac{8m^{3/2}}{15\pi^2\hbar^3}
\left( g_{11} n_1+g_{22} n_2 \right)^{5/2},
\end{equation}
where we assume that atomic masses of both components are equal to $m$.  $n_1$ and $n_2$ are the atomic densities related to the corresponding wave functions $n_i=|\psi_i|^2$. We assume intraspecies repulsion of strength $g_{ii}>0$ and interspecies attraction proportional to $g_{12}<0$, where $g_{ij}=4\pi \hbar^2 a_{ij}/m$ and  $a_{ij}$ are the s-wave scattering lengths. Moreover attraction slightly dominates over repulsion, $\delta g =g_{12}+\sqrt{g_{11}g_{22}}< 0$. The energy functional accounts for quantum fluctuations given by the Lee-Huang-Yang (LHY) term \cite{LHY}. For negative values of $\delta g<0$  this contains a small imaginary part. Second-order Beliaev theory for mixtures allows us to account for higher order terms \cite{Beliaev,Utesov}  which may cure this problem. However, the Beliaev approach for nonuniform two-component mixtures is extremely challenging. Alternatively, a phenomenological approach based on thermodynamic relations was suggested \cite{Ota20}, but sound velocities are still imaginary for very small densities $na_{11}^3 \ll (\delta g/g)^2$ in this framework. Another direction is to include some presumably missing small contributions to the system energy. It was shown in \cite{Hui20} that adding a pairing energy removes the imaginary LHY component. Despite all these different efforts no consistent ultimate solution to the problem of imaginary LHY energy is present yet. For the purpose of the present analysis we simply neglect the imaginary term, as is commonly done in theoretical studies of quantum droplets. This omission is justified close to the instability threshold.

Self-bound droplets are formed for particular values of $N_i=N^0_i$ \cite{Zin20}. The parameter $a=a_{22}/a_{11}$ determines the ratio of atom number in the limit of very large droplets, $N^0_1/N^0_2=\sqrt{a}$. 
We use $n^0_1=\frac{25\pi}{1024}\frac{|\delta a|^2}{a_{11}^5a(1+\sqrt{a})^5}$  to calibrate atomic densities, $n_i= n^0_1 |\Psi_i|^2 $. This way, for large droplets we have $|\Psi_1|^2=1$ and $|\Psi_2|^2=1/\sqrt{a}$ within the bulk. Following \cite{Petrov15}  we set  $\xi=\sqrt{\frac{3}{2}\frac{1+\sqrt{a}}{4\pi |\delta a|n^0_1}}$,  $\mu_0=\hbar^2/(m\xi^2)$, $t_0=\hbar/\mu_0$ as units of length, energy and time respectively. The number of particles in the respective components of the mixture equals $N^0_i = {\cal N}_0  \int d^3 x |\psi_i|^2$, where ${\cal N}_0=n_1^0 \xi^3$. The  total energy will be expressed in "extensive units" $e= \mu_0 {\cal N}_0= n_1^0\xi\hbar^2/m$. In calculations we set $m = 38.96 u$, where $u$ is the atomic mass unit, and assume symmetric interactions, $a_{11} = a_{22} = 33.83 a_B$, where $a_B$ is the Bohr radius, and $\delta a = -0.0664 a_{11}$. This results in $\xi = 1.04\mu$m, $t_0 = 67\mu$s, 
and ${\cal N}_0 = 2097$.

The ground state is obtained by minimization of the following energy functional:
\begin{equation}
   E[\Psi_1,\Psi_2]=\int d{\bf r} ({\cal E} + \sum_i (|\nabla \Psi_i|^2/2  -\mu_i |\Psi_i|^2)), 
\end{equation}
where the interaction energy density equals ${\cal E}=\epsilon (|\Psi_1|^2,|\Psi_2|^2)/\mu_0$.
The two chemical potentials $\mu_i=\mu_i(N_1,N_2)$  are the only free parameters. 
Dynamical extended Gross-Pitaevskii equations (eGP) consistent with the above energy functional have the form:
\begin{equation}
\label{eq:GP2skl}
i \frac{\partial}{\partial t} \Psi_i =  \left[-\frac{1}{2}\nabla^2+
\frac{\delta {\cal E}}{\delta \Psi^*_i}\right]\Psi_i,
\end{equation}
Stationary solutions of Eq.~(\ref{eq:GP2skl}) are $\Psi_i(t)=\Psi_i(0) e^{-i\mu_i t}$.
The eGP equations quite accurately describe large Bose-Bose droplets \cite{Cabrera18,Tarruell18,Semeghini18,G_Ferioli20}. In order to get quantitative agreement for small droplets, more sophisticated approaches are needed \cite{Mazzanti2019}. A beyond local density approximation to the LHY energy is desirable then. 

In our description we do not account for three-body losses, $dN/dt \simeq  - K |\Psi_i|^6$. They may be accounted for by including an appropriate imaginary term on the right hand side of Eq.~(\ref{eq:GP2skl}). As shown in the experiment \cite{Fattori19} the effect of this term was important for qualitative agreement between the experimental results and a theory based on the eGP equation. In our case this term would play a similar role. At present, the life-time of homonuclear droplets ($\sim 10$ms) is shorter than the complete process, i.e. preparation of a suitable initial state and collision. We do not include this effect since our aim is to study which collision scenarios are possible once three-body losses are overcome. We believe that theoretical and experimental progress in the field will allow for such long-lived droplets in the future. Heteronuclear droplets in lower dimension are a possible remedy, but other unexpected solutions can not be ruled out.

\section{Initial state}
\label{sec:3}
Collisions of two droplets were studied experimentally in \cite{Fattori19}. The initial state was prepared by forming two droplets in a double well potential followed by removal of the barrier separating the two wells. No relative phase was imprinted onto the droplets and the results obtained were consistent with the assumption that their phases are identical. Therefore many aspects of the collisions were similar to classical liquid droplets. For low velocities the two droplets merged while for larger velocities they separated after the collision. The critical velocity was found to depend on the droplet size. The effect of size in collisions of 1D droplets was discussed in \cite{BAMalomed18}.

Here we consider dynamics of interacting droplets having initially nonzero relative phases and moving in opposite directions. Such an arrangement can be experimentally achieved in a way similar 
to those of experiment \cite{Fattori19}. After separation however, a desired phases should be optically imprinted  onto the droplets \cite{Andrelczyk01}.  

The preparation of an initial state with well controlled initial phases is a challenging task. In the ideal situation, splitting of a single droplet into two symmetric smaller ones ought to be performed adiabatically in order to avoid excitations, which in the case of a droplet would lead to evaporation of particles. Special attention should be given to the symmetric placing of the splitting barrier. This was successfully achieved in experiments with a quantum gas placed in a double well potential and in Josephson-junction studies \cite{Inguscio2001, Oberthaler2005, Levy2007, Roati2015}. Here, in principle, the same procedure might be implemented

In the following we assume that the initial state is a {\it superposition} of  two-component waves moving towards each other (compare Eq.(\ref{psi_sc})):
\begin{equation}
\label{psi_sc2}
\Psi^{ini}_i({\bf r}) \sim \phi^{ini}_{i,L}({\bf r}) + \phi^{ini}_{i,R}({\bf r})
\end{equation}
 where $i=1,2$ enumerates components, and  $\phi^{ini}_{i,L(R)} = \Phi_{i,L(R)}({\bf r})e^{\pm i({\bf p}{\bf r}-\Delta \phi/2)}$. 
Such an initial state dovetails the $N$-body "superfluid" state:
\begin{equation}
\label{coh}
    \Phi_i({\bf r}_1,\ldots,{\bf r}_N)=\prod_{i=1}^{N} \left( \phi^{ini}_{i,L}({\bf r}_i) + \phi^{ini}_{i,R}({\bf r}_i) \right),
\end{equation}
with every atom being in the superposition of the left and right droplet.  The phase between the right and left droplet is well controlled, i.e. does not vary from one realization of the system to the other. However, the number of atoms 
fluctuates.  It follows from Eq.~(\ref{coh}) that for large $N$ each droplet is in a coherent state and has on average $N/2$ atoms with a dispersion equal to $\sqrt{N}/2$. The one particle density matrix has only one nonzero eigenvalue  and a corresponding eigenvector, Eq.~(\ref{psi_sc2}). This is a situation ideally suited for eGP equations.

On the other hand one may consider a situation where the two droplets are prepared independently. The $N$-body state is then a product of two Fock states:
\begin{equation}
\label{mot}
\Phi_i({\bf r}_1,\ldots,{\bf r}_N)=\sum_P \prod_{i=1}^{N/2}\Psi_L(P{\bf r_i}) \prod_{i=N/2+1}^{N} \Psi_R(P{\bf r_i}),
\end{equation}
where $P$ stands for permutations of the $N$ atom positions. From the point of view of one-particle measurements, the system is fragmented, with two eigenvalues equal to 1/2 each. The two eigenvectors of the one-body density matrix correspond to two wavefunctions: the right and the left one, $|\phi^{ini}_{i,L(R)}({\bf r})|$.
The multiconfiguration time-dependent Hartree approach could be a right tool to tackle such a problem \cite{Alon2006,Alon2008,Alon2014,Alon2020,Alon2021}.

The  fragmentation  of  one-particle  density  is  the  result  of  a  lack  of  a  well defined  relative  phase,  which strongly fluctuates from one realization of the system to the  other. However, in every single  realization (for repeated measurements) it has some (uncontrolled) value which is fixed in course of a measurement. This issue was discussed in detail in the context of interference of two Fock states \cite{Juha,Dragan}. In a real cold-atom experiment the system is monitored by a CCD camera where a shadow of the atomic cloud is captured at some given instant of time. In  such  a  single  shot  many  atoms, $\sim d \gg 1$,  interact with light and are monitored simultaneously. Therefore all observables ought to be averaged with respect to a $d$-body density matrix which in case of the product of two equally populated Fock-states Eq.~ (\ref{mot}) was found in \cite{Dragan}:
\begin{equation}
\label{d-state}
\rho({\bf r}_1,\ldots, {\bf r}_d) =\int_0^{2\pi}\frac{d\phi}{2 \pi}\prod_{i=1}^d \left|\Phi_L({\bf r}_i) +e^{i\phi} \Phi_R({\bf r}_i) \right|^2,  
\end{equation}
where the limit $N \gg d \gg 1$ was  assumed. Thus even if the system seems to be fragmented, its  dynamics in a single realization may be safely described assuming a superposition of the left and right droplet states, Eq.~(\ref{d-state}) with some arbitrary phase $\phi$. The phase will differ from one realization to the other but is fixed for all detected atoms in a single measurement. Therefore it is justified to use eGP equations also in the case of independently prepared droplets. Some information about the actual relative phase can be eventually deduced {\it a posteriori} from the observed scattering dynamics.

\section{Interaction potential}
\label{sec:4}
At large distances the eGP equations simplify to  $-\frac{1}{2}\nabla^2 \Psi_i = \mu_i \Psi_i$, assuming spherical symmetry. The solutions are $ \Psi_i( {\bf r} )=A_i \frac{e^{-\lambda_i r}}{r}$, where $\lambda_i = \sqrt{-2\mu_i}$. Coefficients $A_i$ and $\lambda_i$ depend on the number of particles of the two species in the droplet. To find the interaction potential we follow the approach presented in \cite{Malomed98,Pawlowski15}. 
We assume that the two droplets are separated by a distance significantly larger then their radii. The exponentially vanishing tails of their wavefunctions overlap and  contribute to the interaction energy. The total wavefunctions $\Psi^{Sc}_i$ of the system are assumed to be sums of two stationary droplet solutions, a left and right one, having different phases: $\Delta \phi_1=\phi^R_1-\phi^L_1$ and  $\Delta \phi_2=\phi^R_2-\phi^L_2$, and separated by a distance $R=|{\bf r}_R- {\bf r}_L|$:  
\begin{equation}
\label{psi_sc}
\Psi^{Sc}_i({\bf r})=\sqrt{{\cal B}_i} \sum_{\alpha=L,R}\Psi_i( {\bf r}-{\bf r}_\alpha )e^{-i{\bf p}_\alpha{\bf r}-i\phi^\alpha_i},
\end{equation}
where  ${\cal B}_i$ are chosen to ensure normalization of the wavefunction $\Psi^{Sc}_i$ to the value of $(N^0_i)^L+(N^0_i)^R$. 

The interaction energy of the left and right droplet may be defined as the difference between the energy of two overlapping and two infinitely separated droplets $V(R, \Delta \phi_1,\Delta \phi_2)= E[\Psi^{Sc}_1, \Psi^{Sc}_2] - \sum_{\alpha=R,L} E[\Psi^\alpha_1, \Psi^\alpha_2]$:
\begin{equation}
\label{interaction}
V(R,\Delta \phi_1,\Delta \phi_1)= -\sum_i A_i^2\left(\frac{4\pi}{R}\right) e^{-\lambda_i R} \cos(\Delta \phi_i),
\end{equation}
 The potential depends not only on distance $U_i(R) = A_i^2(\frac{4\pi}{R}) e^{-\lambda_i R}$, but also on the phase difference between the droplets, and may be attractive or repulsive. The characteristic range of the interaction is $1/\lambda_i$.
 
The coefficients $A_i$ are found from the continuity condition, $|\Psi_i(R_0)| = A_i  \frac{e^{-\lambda_i R_0}}{R_0}$ where $R_0$ is the radius of a droplet having $N^0_i$ atoms, $\left(\frac{4\pi R_0^3}{3}\right) |\Psi_i|^2  = N^0_i$. This is an approximate expression valid only for very large droplets, where the bulk density may be approximated by  $n_{1(2)}=|\Psi_{1(2)}|^2=1, \left(\frac{1}{\sqrt{a}}\right)$. 
In the general case, the coefficients $A_i$ are to be determined by fitting to a numerical solution. This is the approach of the present study.

Equation (\ref{interaction}) is valid for two droplets with equal number of atoms. For a more general treatment, we compare this analytic result with numerical calculations of scenarios in which there is a small population imbalance between the droplets, to obtain the following formula for the spatial part of the interaction potential:
\begin{equation}
    U_i(R)=\frac{4\pi}{R}A^{(L)}_i A^{(R)}_i e^{-(\lambda^{(L)}_i + \lambda_i^{(R)})R/2},
\end{equation}
 The exponents $\lambda^{L,R}_i$ characterize the exponential tails of the wavefunctions $\Psi^{L,R}_i$.

\section{Two-component Josephson junction equations}
\label{sec:5}
Binary droplet collisions may be described by a set of time-dependent 3D partial differential eGP equations - Eqs.~(\ref{eq:GP2skl}). Initially the wavefunction is a sum of two stationary solutions $\Psi^{Sc}_i({\bf r})$. If the initial kinetic energy of translational motion of the droplets is low we may assume that they move as a whole without internal excitations, preserving their shape.  As long as their relative separation is larger then the diameter $2 R_0$ of a droplet, i.e. if $|{\bf r}_{L}(t)-{\bf r}_{R}(t)| > 2R_0$ we may assume that no Bogoliubov quasiparticles are excited. The exponential density tails overlap forming a weak link. The phase difference between the left and right parts of the two wavefunctions will trigger a coherent flow of atoms between the droplets. These are the Josephson-junction-like oscillations of particle number and relative phase. 

Instead of solving the full set of eGP equations, Eq.~(\ref{eq:GP2skl}), the oscillations may be described adequately by only considering the modes involved in the dynamics -- these are the zero-energy or Goldstone modes \cite{Dziarmaga04,Zin20zero}. The energy of a free droplet does not depend on a particular choice of phases of its wavefunctions nor on its position in space. The mean field solutions break these continuous symmetries. In consequence, zero-energy modes which recover the broken symmetries appear in the excitation spectrum. Dynamics of zero modes in a two droplet system is given by the Hamiltonian \cite{Zin20zero}: 
\begin{eqnarray}
\label{HG}
{\cal H}&=&\sum_{\alpha} \frac{{\bf p}_\alpha^2}{2M^\alpha} +
V(| {\bf r}_L - {\bf r}_R |, \Delta \phi_1, \Delta \phi_2) \nonumber \\
&& + \sum_{\alpha}\left( \frac{(p_H^{\alpha})^2}{2 M_H^\alpha}+
\frac{(p_S^{\alpha})^2}{2 M_S^\alpha}+\sum_{i=1,2}\mu_i^\alpha \delta N^\alpha_i\right).
\end{eqnarray}
In above ${\bf p}_\alpha= -i\hbar\frac{\partial}{\partial {\bf r}_\alpha}$ are kinetic momenta of left and right droplet,  $\alpha=L,R$, while ${\bf r}_{\alpha}$ are positions of droplets' centres.  Kinetic momenta of hard and soft modes, $p_H^{\alpha}$ and $p_S^{\alpha}$, are 
\begin{equation}
   p_{S,H}^\alpha = \frac{1}{\sqrt{2}}\left((\mu_{1,1}^\alpha)^{1/2}\delta N_{1}^{\alpha} \pm (\mu_{2,2}^\alpha)^{1/2}\delta N_{2}^\alpha\right),
\end{equation}
where:
\begin{equation}
    \mu_{i,j}^\alpha = \frac{\partial \mu_i^\alpha}{ \partial N_j^\alpha},
\end{equation}
and  the plus (minus) sign refers to the soft (hard) mode respectively while $\delta N_i^{\alpha}$: 
\begin{equation}
 \delta N_i^{L(R)}= N_i^{L(R)}-(N^0_i)^{L(R)},   
\end{equation}
are deviations of particle numbers $N_i^\alpha$ from their equilibrium values. The pairs $({\bf r}_\alpha, {\bf p}_\alpha)$  and  $(\phi_i^\alpha, \delta N_i^\alpha)$ are three sets of canonically conjugate variables.  Coefficients $M^\alpha$   are masses of the two droplets,  $M^{\alpha}=(N_1^0)^{\alpha}+(N_2^0)^{\alpha}$ while  $M_{H,S}^\alpha$ are "masses" of the two phase-modes:
\begin{equation}
 \frac{1}{M_{S(H)}^\alpha}=\left(1 \pm \frac{\mu_{1,2}}{\sqrt{\mu_{1,1} \mu_{2,2}}}\right).   
\end{equation}
The upper sign "+" corresponds to the mass of the soft mode while the lower sign "-" represents the hard mode. As shown in \cite{Zin20zero} the values of $M_S$  and $M_H$ differ significantly, $1/M_H^\alpha \sim 1  \gg  1/|M_S^\alpha| \propto |\delta a|/\sqrt{a_{11}a_{22}}$. This inequality justifies addressing the corresponding excitations as hard and soft modes. In addition, the mass of the soft mode is negative  and changes its sign only for very large droplets. ${\cal H}$ may be treated as a quantum Hamiltonian \cite{Dziarmaga04,Sacha09,Mochol12,Plodzien12}, however for the present purpose we restrict ourselves to a classical treatment. This means neglecting quantum fluctuations of $\phi^\alpha_i$ and $\delta N^\alpha_i$. The equations of motion generated by ${\cal H}$ ensure conservation of particle number of each species and the center of mass momentum. It is convenient to use relative coordinates, ${\bf R}={\bf r}_R-{\bf r}_L$ and  ${\bf P}={\bf p}_R-{\bf p}_L$.  Furthermore we assume that initial orbital angular momentum is equal to zero and the droplets move towards each other along the x-axis, so we may omit vector notation. The equations of motion for the relevant quantities are:
\begin{eqnarray}
\label{NewtonP}
&&\dot{ P}             =  U^{\prime}_1(R) \cos(\Delta \phi_1) + U^{\prime}_2(R) \cos(\Delta \phi_2),\\
\label{NewtonR}
&&\dot{R}             =  \frac{P}{{\cal M}},\\
\label{NewtonN}
&&\delta \dot{N}^R_i  = - \delta \dot{N}^L_i =  -U_i(R) \sin(\Delta \phi_i),\\
\label{Newtonphi}
&&\Delta \dot{\phi}_i  = \Delta \mu_i+ \left(\mu^R_{i,1}\delta N^R_1-\mu^L_{i,1}\delta N^L_1\right) \nonumber \\
&&+\left(\mu^R_{i,2}\delta N^R_2-\mu^L_{i,2}\delta N^L_2\right),
\end{eqnarray}
where ${\cal M}=\frac{M_LM_R}{M_L+M_R}$ is the reduced mass and $\Delta \mu_i=\mu^R_i-\mu^L_i$. Dots denote time derivatives while primes denote spatial derivatives. These are the Josephson junction (JJ) equations for a two-component junction. Relative phases are coupled to currents Eq.~(\ref{NewtonN}) and to the relative motion Eq.~(\ref{NewtonP}). The two Josephson currents $\delta \dot{N}^R_1$, $\delta \dot{N}^R_2$ are mutually coupled via the relative phases $\Delta \phi_i$. This coupling  signifies the Andreev-Bashkin (AB) effect \cite{Nespolo17,Parisi18,Syrwid21,Blomquist21}.

\section{Coherent dynamics}
\label{sec:6}
In our simulations, the centers of droplets are separated by the distance $|{\bf r}_R-{\bf r}_L|=r(0) \simeq 15$.  Due to their finite size, the distance between their surfaces is smaller,  $r(0)-2 R_0 \approx 5-10$. This is comparable to the range of the potential $d=1/\sqrt{2 \mu_1} \simeq 1/\sqrt{2 \mu_2} \approx 1.2-2.2$. 
Initial droplets' velocities are zero ensuring low energy collision. Droplets' densities do not depart significantly from the equilibrium values. Any deviations result in a weak atomic evaporation. Fragmentation of droplets as observed in \cite{Fattori19} occurs only at higher collision energy. In this regime more sophisticated approaches, like multiconfigurational time-dependent Hartree method 
\cite{Alon2021}, might be beneficial.

We begin our discussion by considering the case where both droplets have the same number of atoms and the phases of both components of a droplet are identical,  $\Delta \phi \equiv  \Delta \phi_1 = \Delta \phi_2$. In such an arrangement the system can be described by a single wavefunction. 

Our interpretation is based on the JJ equations. If the droplets are identical and initially $|\Delta \phi_1\pm\Delta \phi_2|<\pi$, they attract mutually and begin to move towards each other. The DC Josepshon current of atoms $\delta \dot{N}_i \sim \sin(\Delta \phi)$ flows from one droplet to the other in direction of the phase gradient. The droplets may merge forming an excited droplet. If the initial phases are such that interaction is repulsive, the droplets repel and move away from each other. The JJ current decreases in time as the coupling between droplets gets weaker. This simplistic description already shows that the motion of droplets may strongly depend on the relative phases of their wavefunctions. 

However, the scenario given above is still oversimplified. Attraction between droplets can change to a repulsion. If the direct (DC) Josephson current triggered by a phase difference is large, one droplet may grow at the expense of the other and a difference between their chemical potentials will develop. This condition supports an alternating  Josephson current (AC) rather then a DC one. The character of JJ dynamics switches to AC mode when the phase difference grows linearly in time, $\Delta \phi \sim \omega t$. The slow motion of droplets and fast phase dynamics happen on two different time scales. Averaging over fast oscillations according to P.L.~Kapitza's method \cite{Landau} allows to separate a fast micromotion from a slow relative movement governed by a repulsive ponderomotive  potential 
\begin{equation}
V_{pon} \approx \sum_i \frac{U^\prime_i(R)^2}{ 4M\omega^2}.    
\end{equation}
This is a mechanism which leads an initially attractive potential eventually to act repulsively and allows the droplets to escape. 

Due to lack of confinement the distance between the droplets may change, thus the coupling strength varies in time. The Josephson current is a transient effect, and occurs only when the droplets are close to each other. This is a substantial difference from the case of two trapped BECs \cite{Levy2007}.

The Kapitza mechanism is illustrated in the top panel of Fig. 1, showing the interaction of two droplets with no initial phase difference, but with different atom numbers (and therefore different chemical potentials), $N_1^R=N_2^R=20.3 (\times {\cal N}_0) =20.3 (\times 2097)$ and $N_1^L=N_2^L=25.4(\times 2097)$. The AC Josephson effect leads eventually to repulsion, despite initial attraction. The left part of the panel  shows a 1D cut in density along the axis of the droplets' motion (vertical axis in the figure) as a function of time. The density was obtained by numerical solution of the eGP equations.  Black lines in the figure show the droplets' trajectories as obtained from the JJ equations. In the right part of the panel we show: a) the phase difference between droplets $\Delta \phi$ as obtained from solving the eGP equations (taken at maximum density) -- red line,  and b) the difference in atom number $N^R_1-N^L_1$ -- black line. Dashed lines of the same  colors are corresponding results obtained from the JJ equations. A small amplitude micro-motion is visible in the droplets' trajectories. The dashed lines generally coincide with the solid lines and are barely visible. 

 \begin{figure}[htb]     
 \centering
         \includegraphics[trim={1.cm 6.75cm 0.75cm 0},clip,width=0.49\columnwidth]{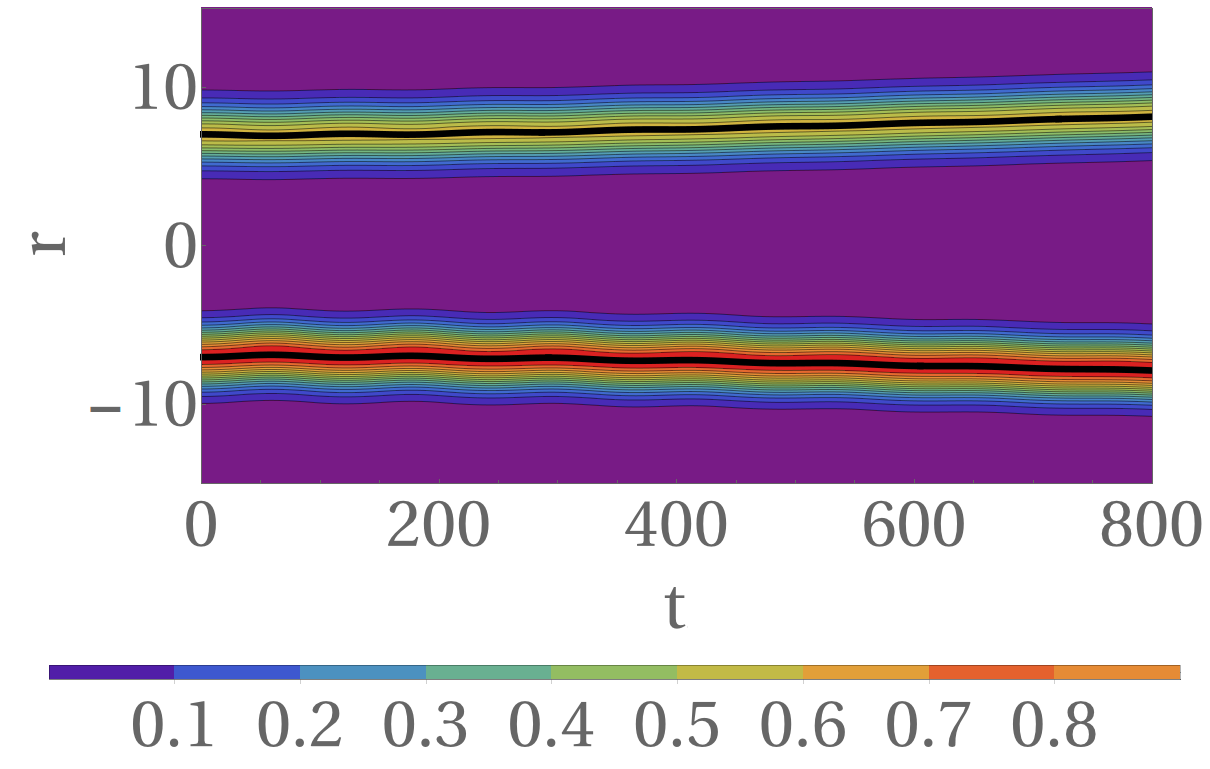}
          \includegraphics[trim={0.75cm 2.75cm 0 0},clip,width=0.49\columnwidth]{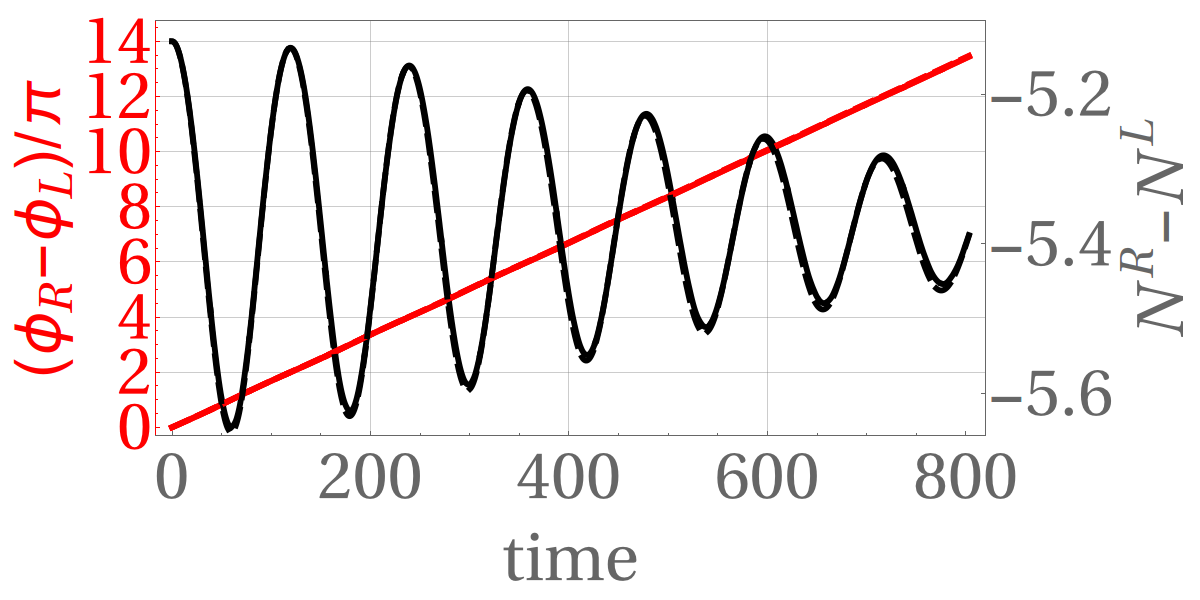}\\
        \includegraphics[trim={1.cm 6.75cm 0.25cm 0},clip,width=0.49\columnwidth]{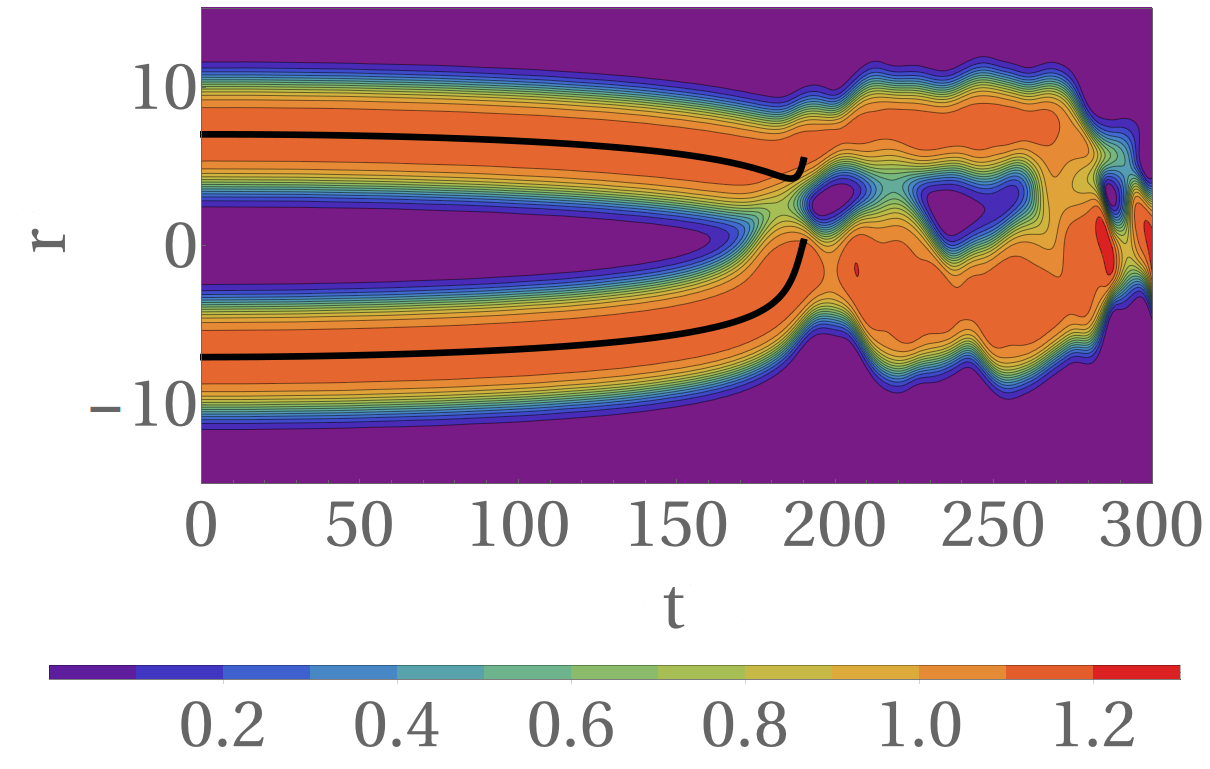}
          \includegraphics[trim={0.75cm 2.75cm 0 0},clip,width=0.49\columnwidth]{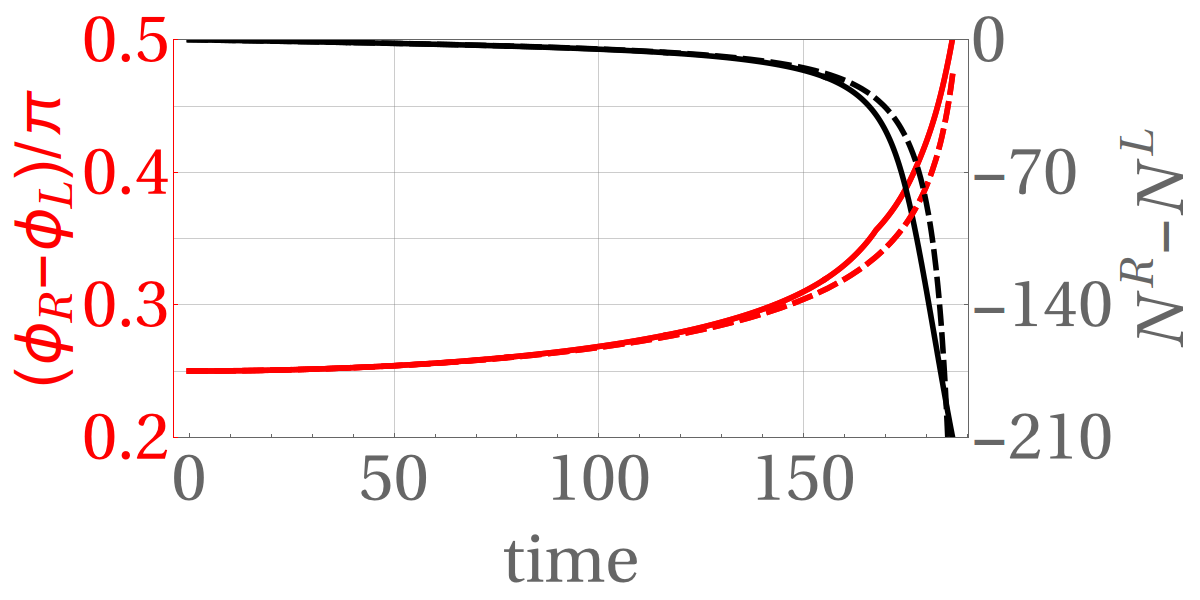}\\
         \includegraphics[trim={1.cm 4.5cm 0.25cm 0},clip,width=0.49\columnwidth]{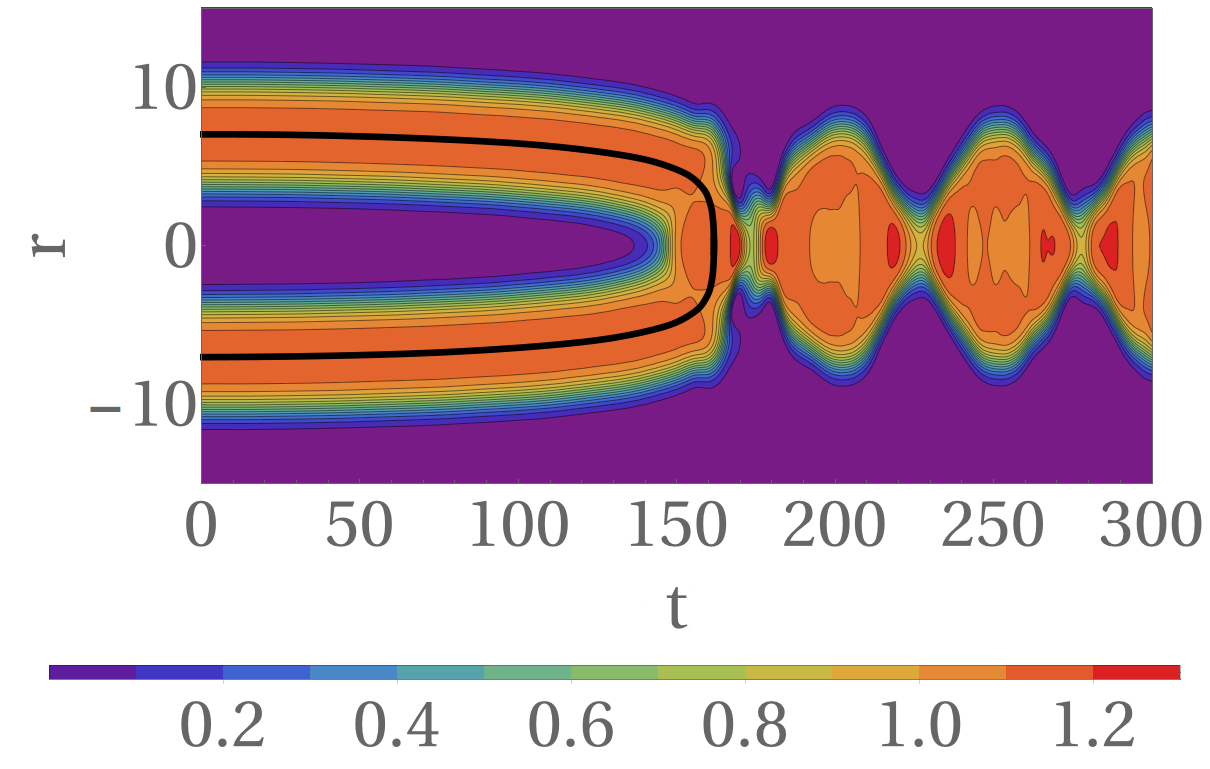}
          \includegraphics[trim={0.75cm -0.5cm 0 0},clip,width=0.49\columnwidth]{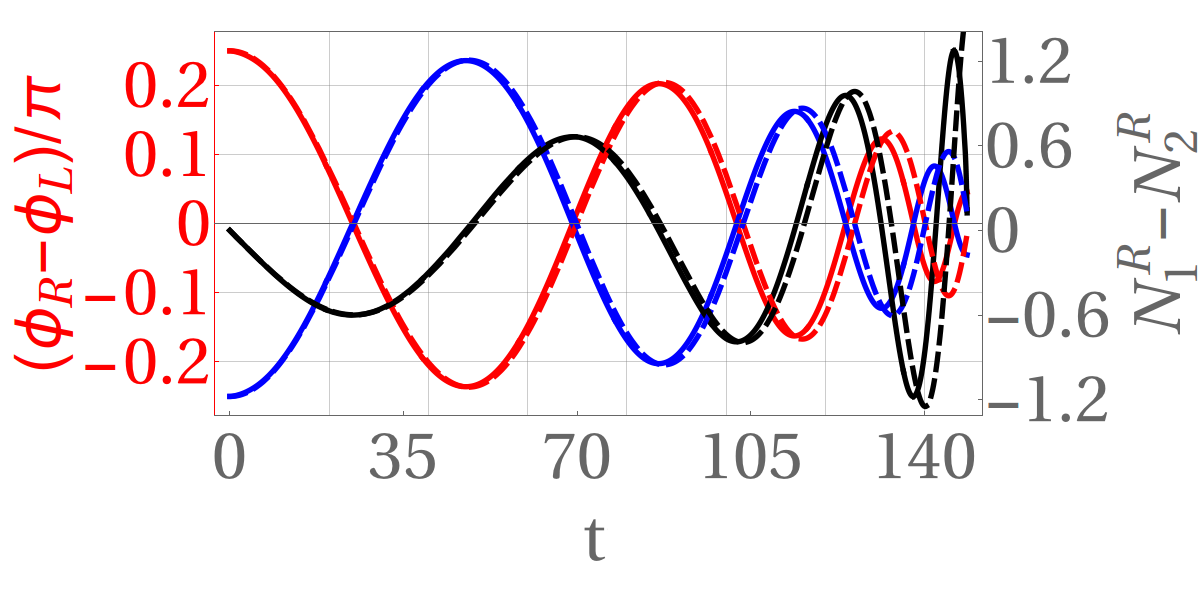}\\
     \caption{Two droplet collisions. Left column: Cut of droplet density along propagation axis resulting from eGP dynamics and droplet trajectories obtained from JJ equations (black lines). Right column: relative phase $\Delta \phi \equiv \Delta \phi_1 = \Delta \phi_2$ (red solid and dashed line) and  number of atoms  $N^R_1-N^L_1 (=N^R_2-N^L_2)$ (black solid and dashed line). Initial conditions: Top: $N^L_1=25.4$, $N^R_1=20.3$,  $\Delta \phi= 0$, middle: $N^L_1=N^R_1=203$,  $\Delta \phi= \pi/4$, bottom: $N^L_1=N^R_1=203$,  $\Delta \phi_1=-\Delta\phi_2=\pi/4$. Two relative phases $\Delta \phi_1$ and   $\Delta \phi_2$ are plotted (red and blue solid and dashed lines respectively).   
     Time is expressed in units of $t_0=0.669$ ms. Difference between results based on JJ equations (dashed lines) and solutions of eGP equations (solid lines) is not visible. Video links: \cite{Csmall_0_0} (top), \cite{Alarge_45_45} (middle), \cite{Blarge_45_-45} (bottom) }
      \label{fig1}
  \end{figure}

 \begin{figure}[htb]     
 \centering
           \includegraphics[trim={1.0cm 6.75cm 0.25cm 0},clip,width=0.49\columnwidth]{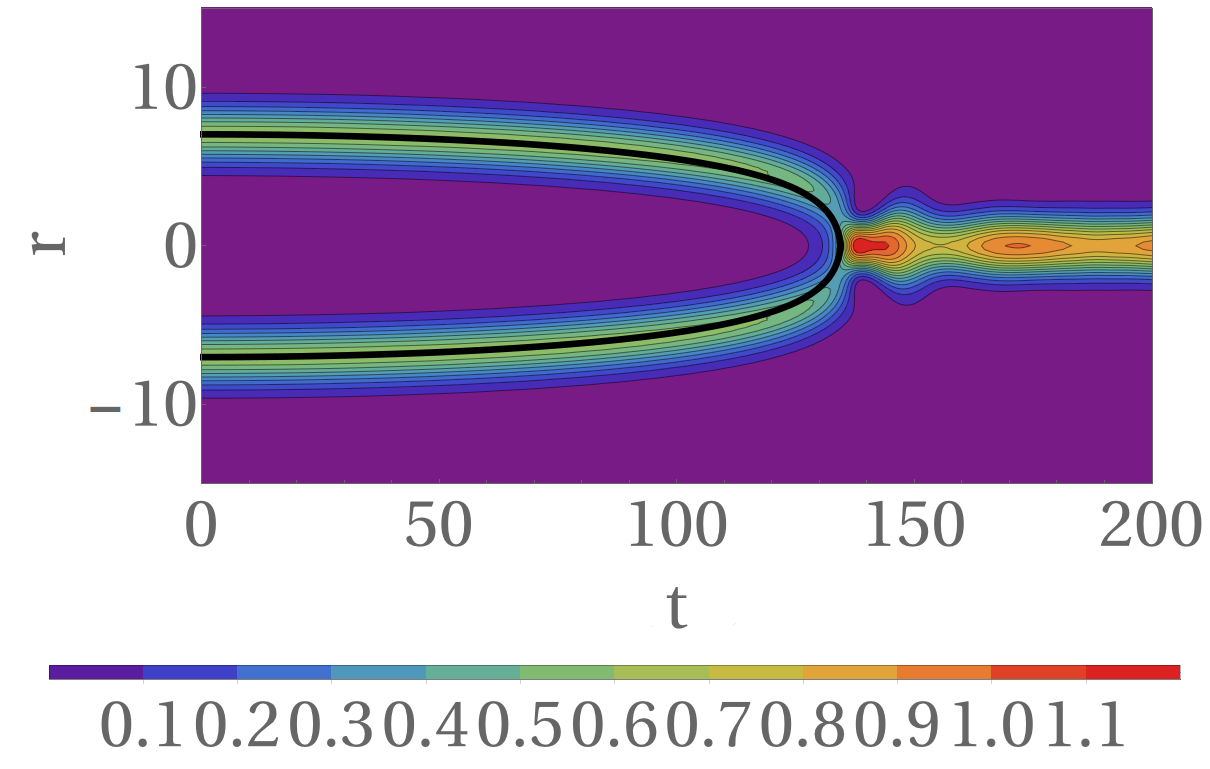}
          \includegraphics[trim={0.75cm 2.75cm 0 0},clip,width=0.49\columnwidth]{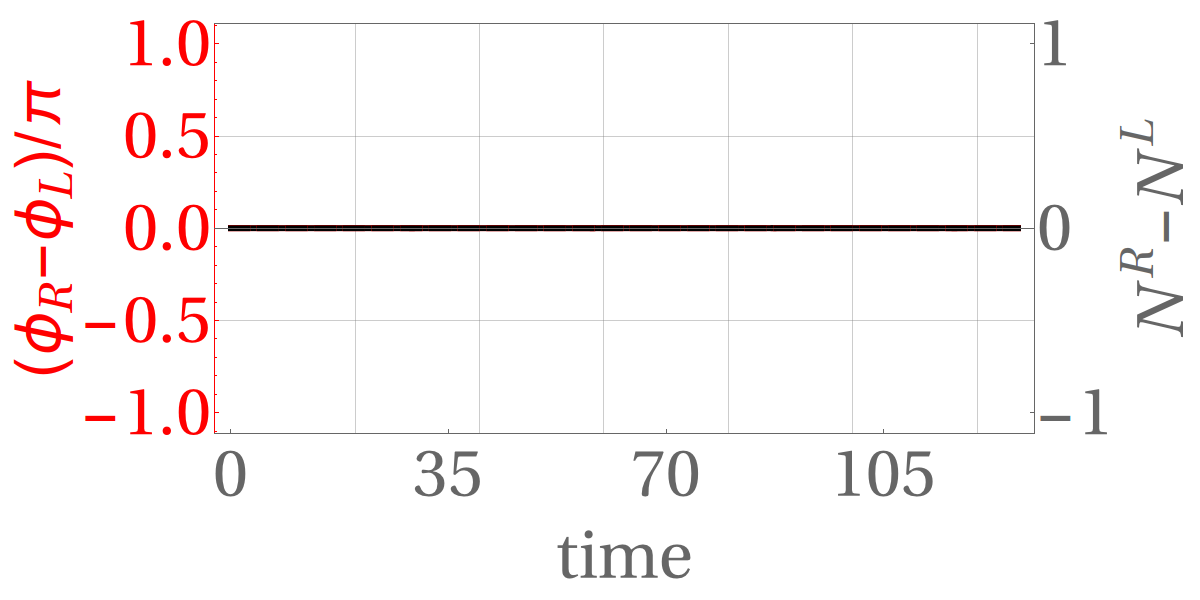}\\
          \includegraphics[trim={1.0cm 6.75cm 0.25cm 0},clip,width=0.49\columnwidth]{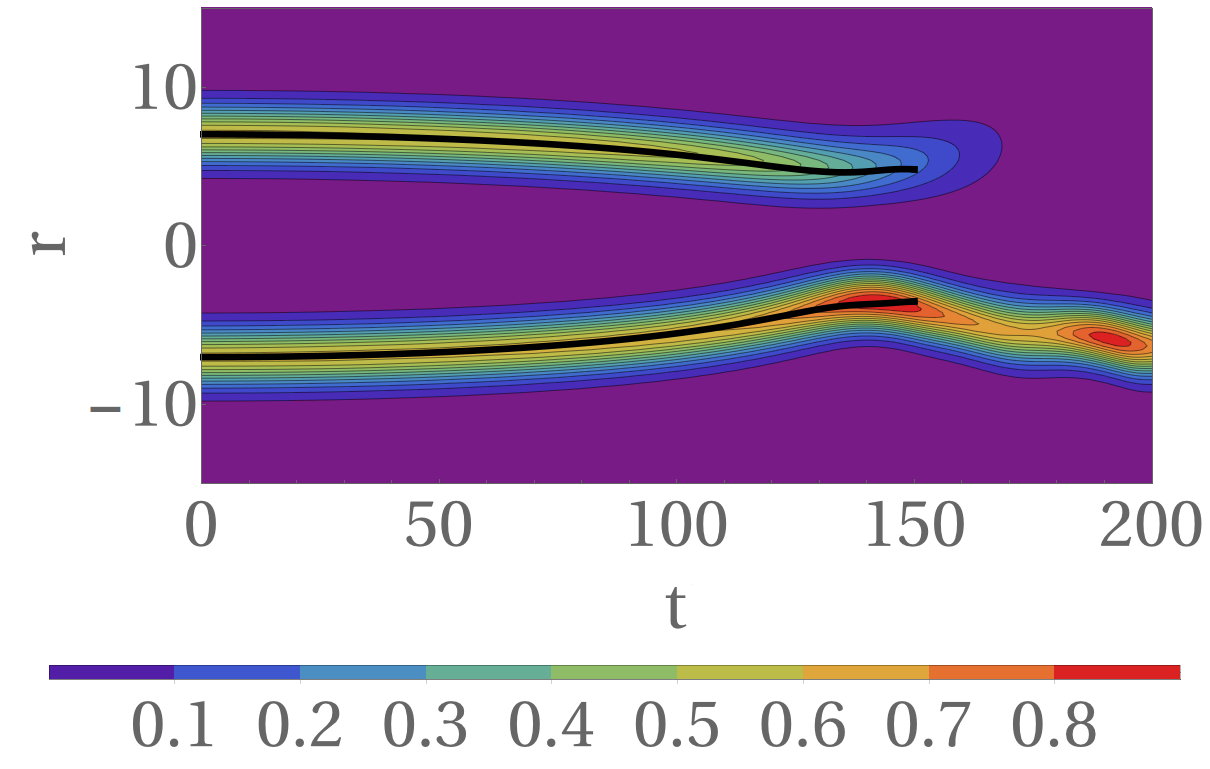}
          \includegraphics[trim={0.75cm 2.75cm 0 0},clip,width=0.49\columnwidth]{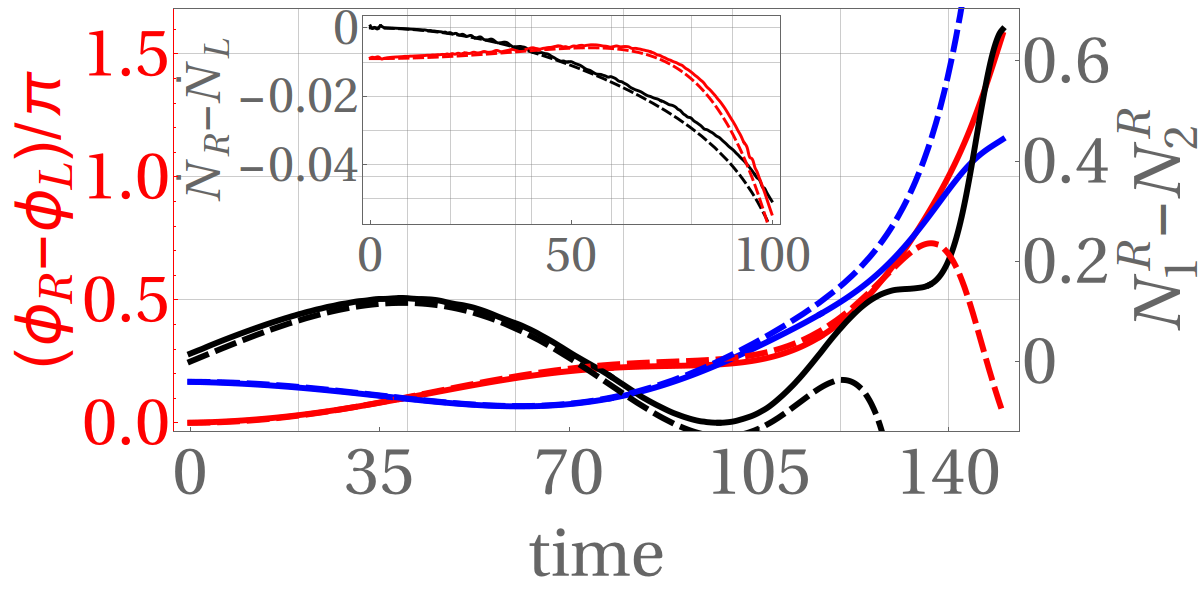}\\
          \includegraphics[trim={1.cm 4.75cm 0.25cm 0},clip,width=0.49\columnwidth]{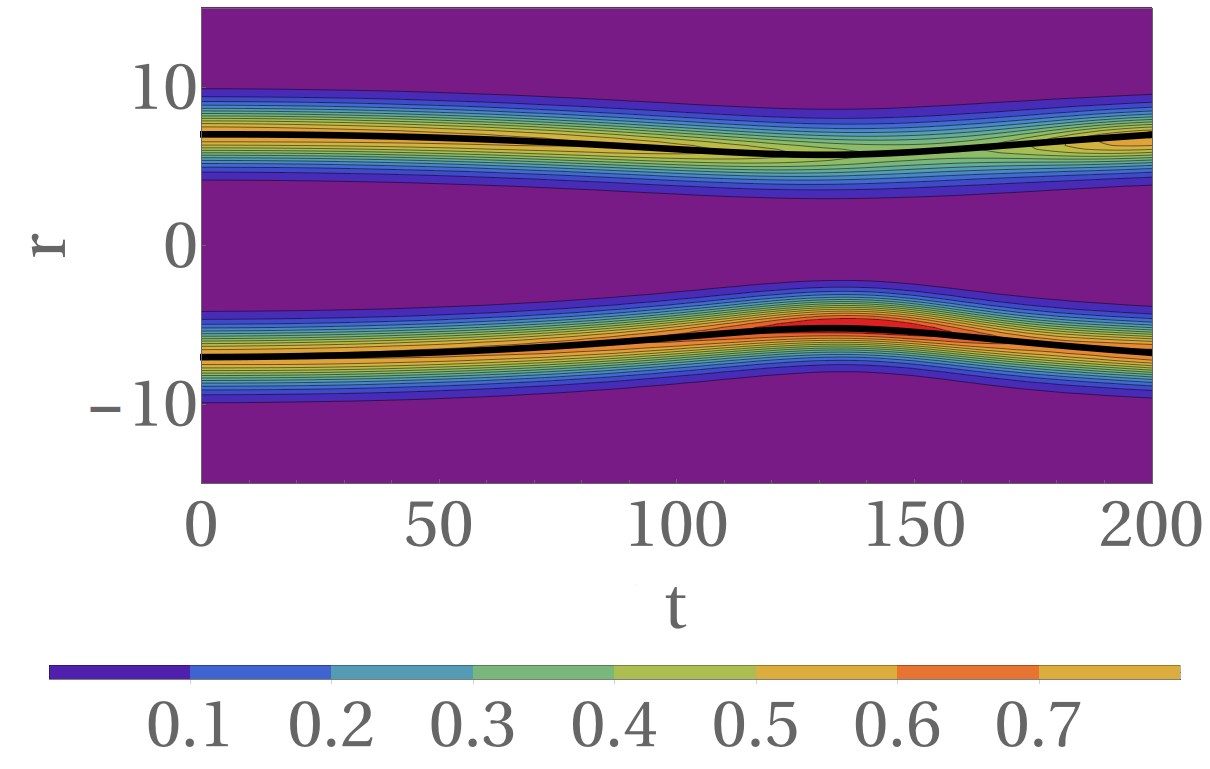}
          \includegraphics[trim={0.75cm 0.5cm 0 0},clip,width=0.49\columnwidth]{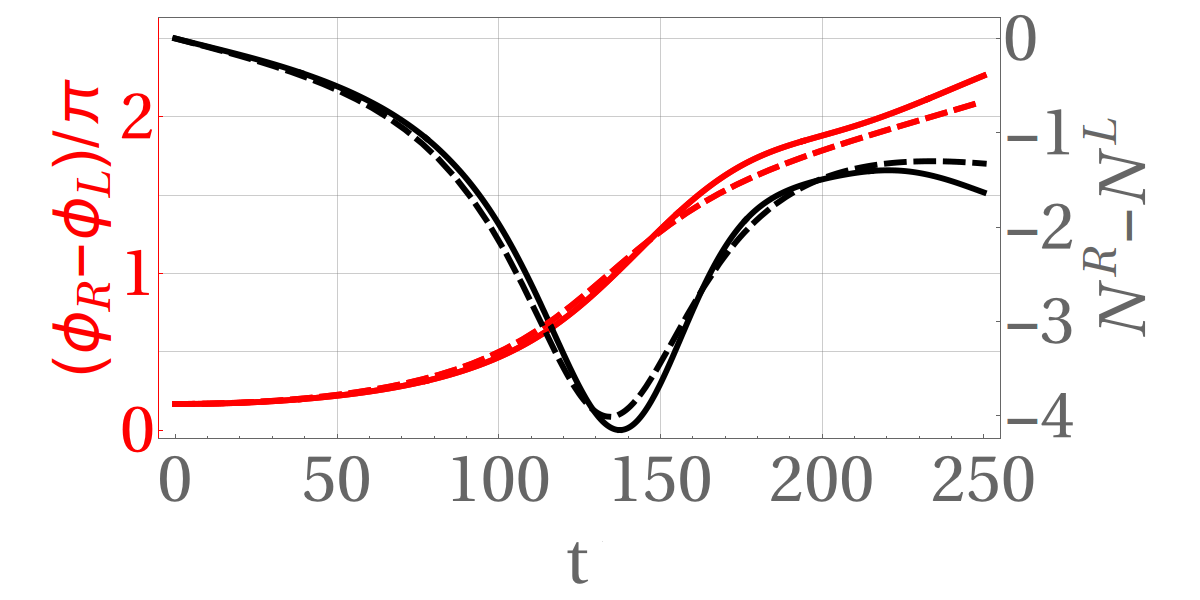}
     \caption{Two droplet collisions. Initial conditions: Top: $N^L_1=N^R_1=20.3$, $\Delta \phi_1=\Delta \phi_2=0$; middle: $N^L_1=N^R_1=20.3$, $\Delta \phi_1=0$, $\Delta \phi_2=\pi/6$; 
     bottom: $\Delta \phi_1=\Delta \phi_2=\pi/6$ (final number of particles $N^R_1=19.48$, $N^L_1=21.12$). Other parameters, dimension of axes and colour coding is the same as in Fig.1. Video links: \cite{Hsmall_0_0} (top), \cite{Ismall_0_30} (middle), \cite{Esmall_30_30} (bottom)}
      \label{fig2}
  \end{figure}

Another situation in which an initial DC current turns into an AC one is illustrated in Fig.1, middle panel, where we show a collision of two large, initially attractive droplets, $N_1^R=N_1^L=203 (\times 2097)$ and $\Delta \phi=\pi/4$. As the droplets approach each other, the "down" one grows as it is continuously supplied by the DC current, while the "up" one becomes significantly smaller, until the AC mode takes over and the Kapitza mechanism comes into play. The smaller droplet is repelled, but before escaping it is recaptured due to the large amplitude of quadruple oscillations of the excited left droplet. This is an example of a very spectacular scenario where one of the droplets steals atoms from the other "at a distance" and finally devours its smaller companion.

When one of the phase gradients has opposite sign, as in Fig.1, bottom panel ($\Delta \phi_1=-\Delta \phi_2 = \pi/4$), the droplets attract each other, and two equal but opposite Josephson currents are initiated. Such a scenario cannot be described by the single-component eGP equation. The two counter-flows oscillate around zero while keeping opposite directions. These oscillations have a tendency to separate both species and accumulate them in opposite droplets. Excess atoms gather mostly in low density regions where some deviation from the equilibrium proportion does not destabilize the droplet \cite{Zin20}. None of the droplets is depleted and eventually they merge, similarly as if they had zero initial phase. For comparison see Fig.2, top panel, where merging of two small droplets, $N_1^R=N_1^L=20.3 (\times 2097)$ is illustrated. Here however, no Josephson current nor phase dynamics take place. The resulting droplet is not as excited as the one discussed previously.

In the middle panel we show the Andreev-Bashkin effect, i.e. entertainment between the two superfluids. We simulate two small droplets $N_1^R=N_1^L=20.3 (\times 2097)$ with vanishing initial phase gradient  $\Delta \phi_1=0$ in one component and $\Delta \phi_2=\pi/6$ in the second one. If the currents were independent, only one species of atoms would flow from one droplet to the other. Yet both species begin to travel together and the related supercurrents are of the DC type. Eventually the "up" droplet loses so many atoms that it becomes unstable and evaporates during the approach.  The minimal number of atoms supporting a stable droplet is about $18.65 \times  {\cal N}_0$. The right droplet disappeared altogether while approaching its partner because both currents were acting in the same direction. The JJ equations do not account for evaporation  and thus are not accurate at the final stage of such dynamics. The inset (middle panel, right) shows the entertainment between the two supercurrents $\delta \dot{N}^R_1-\delta \dot{N}^L_1$ -- black line, and $\delta \dot{N}^R_2-\delta \dot{N}^L_2$  -- red line.  Note that the "black" current starts from zero but soon follows the "red" current. We stress that such a spectacular disappearance of one droplet in the presence of another is not unique and does not necessarily signify the AB effect. Very similar behaviour (not illustrated here) may be observed if both initial phase gradients are equal (e.g. $\Delta \phi_1=\Delta \phi_2=\pi/10$), i.e. if the two supercurrents act together from the very beginning. Therefore experimental proof of the AB effect  requires precise control over the droplets' phases.  
In turn, if both initial phase gradients are equal to $\Delta \phi = \pi/6$, the JJ currents are larger than in the aforementioned case (Fig. 2, bottom panel), droplets initially approach but the AC sets on while the droplets are still far away from each other. Thus repulsion takes over before the smaller droplet disappears. 

Experimental realizations of phase dependent collisions crucially depend on the sensitivity of the entire process with respect to the stability of the relative phase difference. To investigate this problem we randomly perturb the initial phases of the droplets. After preparing the initial state with a well defined phase difference, we add to both components some local phase perturbations:
\begin{equation}
    \Psi^{sc}_{i,L(R)}({\bf r}) \to \Psi^{sc}_{i,L(R)}({\bf r})
    \left( 1+e^{i\delta \phi_{i,L(R)}(\bf{r})}\right),
\end{equation}
where $\delta \phi_{i,L(R)}(\bf{r})$ is a random variable uniformly distributed in the interval $[-\delta ,\delta]$. In the example we show below, we chose the initial phase difference $\Delta \phi_0 = 3\pi/4$  and $\delta = 0.1 \times \Delta \phi_0$. The perturbation is thus quite strong, at the level of $\pm 10\%$.

\begin{figure}[htb]     
 \centering
           \includegraphics[width=0.49\columnwidth]{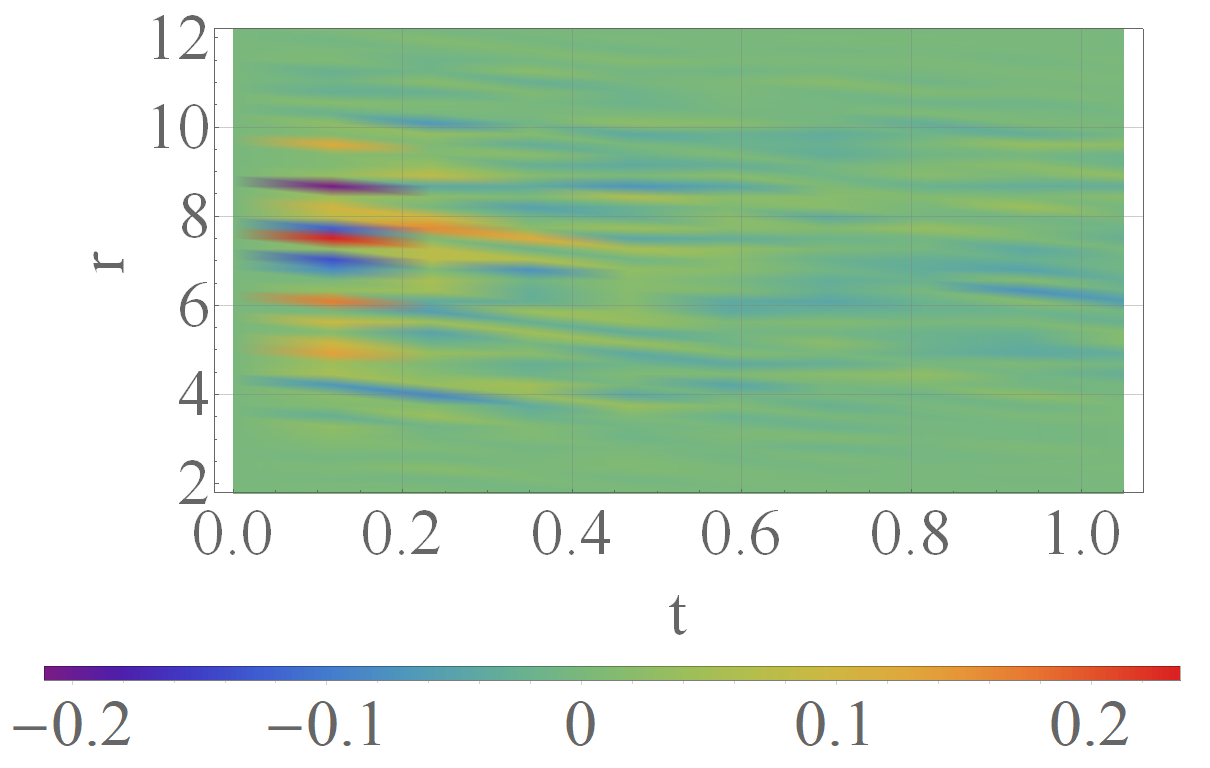}
           \includegraphics[width=0.49\columnwidth]{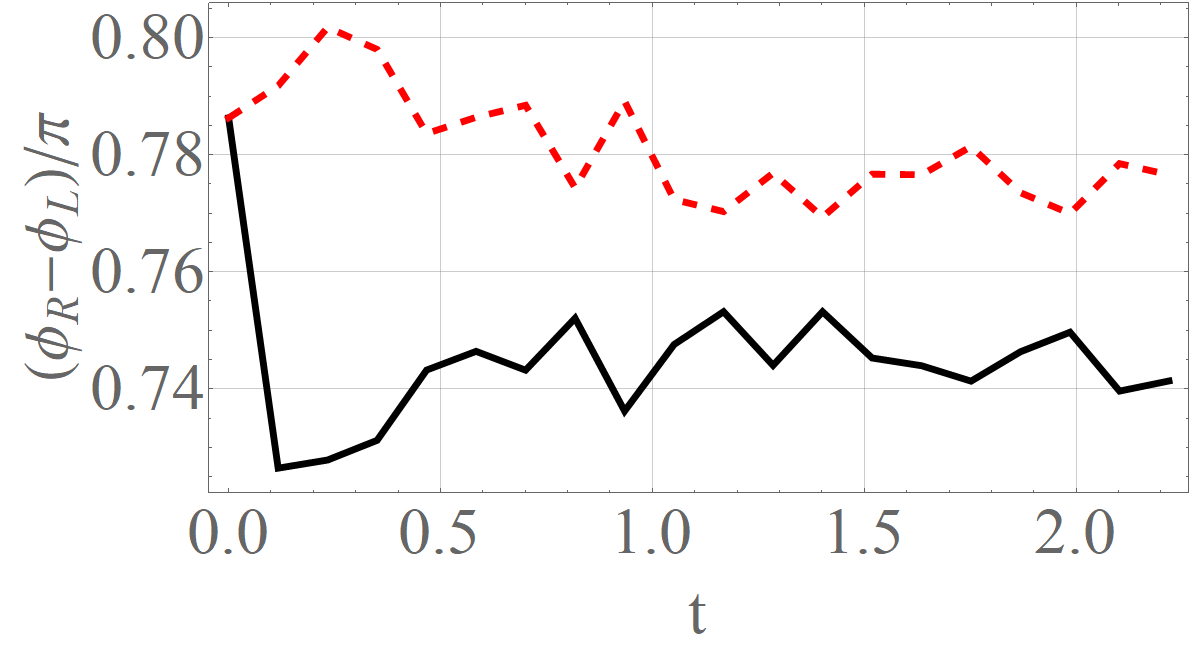}
     \caption{Initial stages of evolution of  the right droplet (located at t=0 at $r_R(0)=7.5$) after introducing phase  perturbations. Left panel:  relative difference of densities of perturbed and unperturbed  droplets. Right panel: difference of phases between the right and the left droplet calculated at positions of their centers  for the first (black solid line) and the second (red dashed line) component. Note that large density fluctuations appear at short time and then about $t=1.0 (0.7ms)$ the phases continue to fluctuate with relatively small amplitudes. }
      \label{dens_diff}
  \end{figure}

\begin{figure}[htb]     
 \centering
           \includegraphics[trim={1.cm 4.75cm 0.25cm 0},clip,width=0.49\columnwidth]{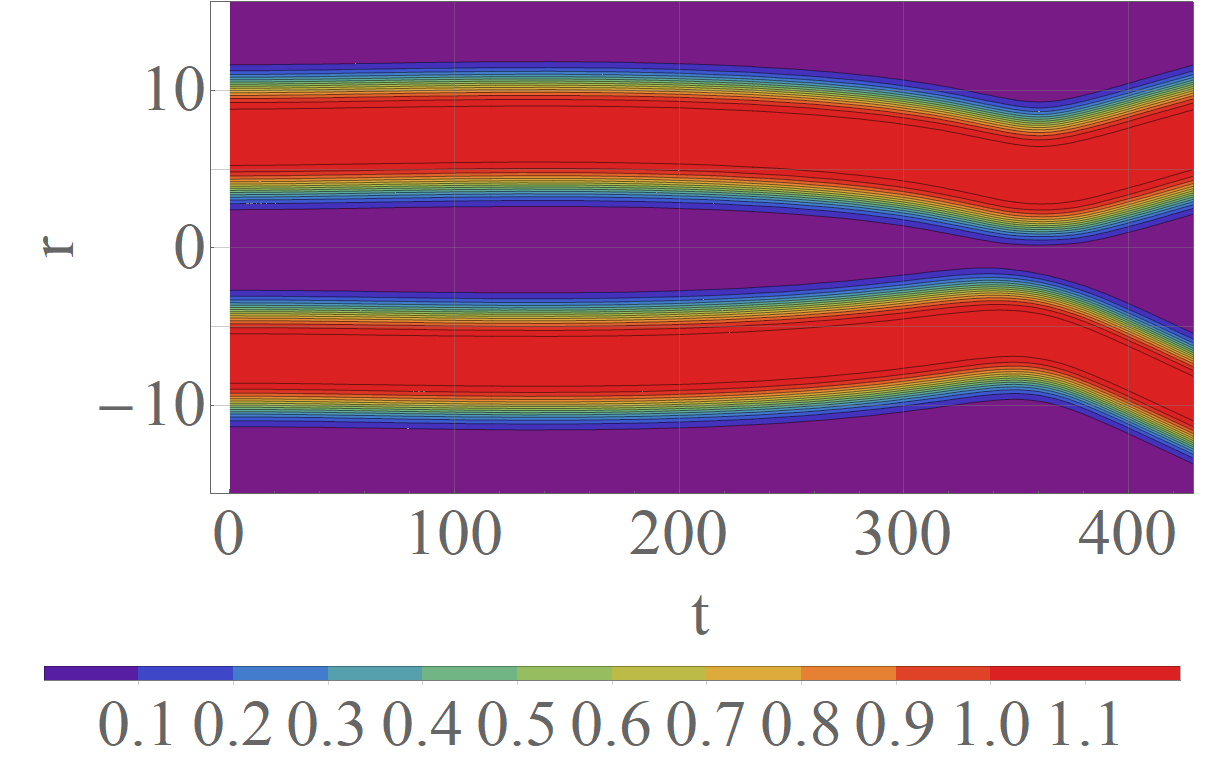}
           \includegraphics[trim={1.cm 4.75cm 0.25cm 0},clip,width=0.49\columnwidth]{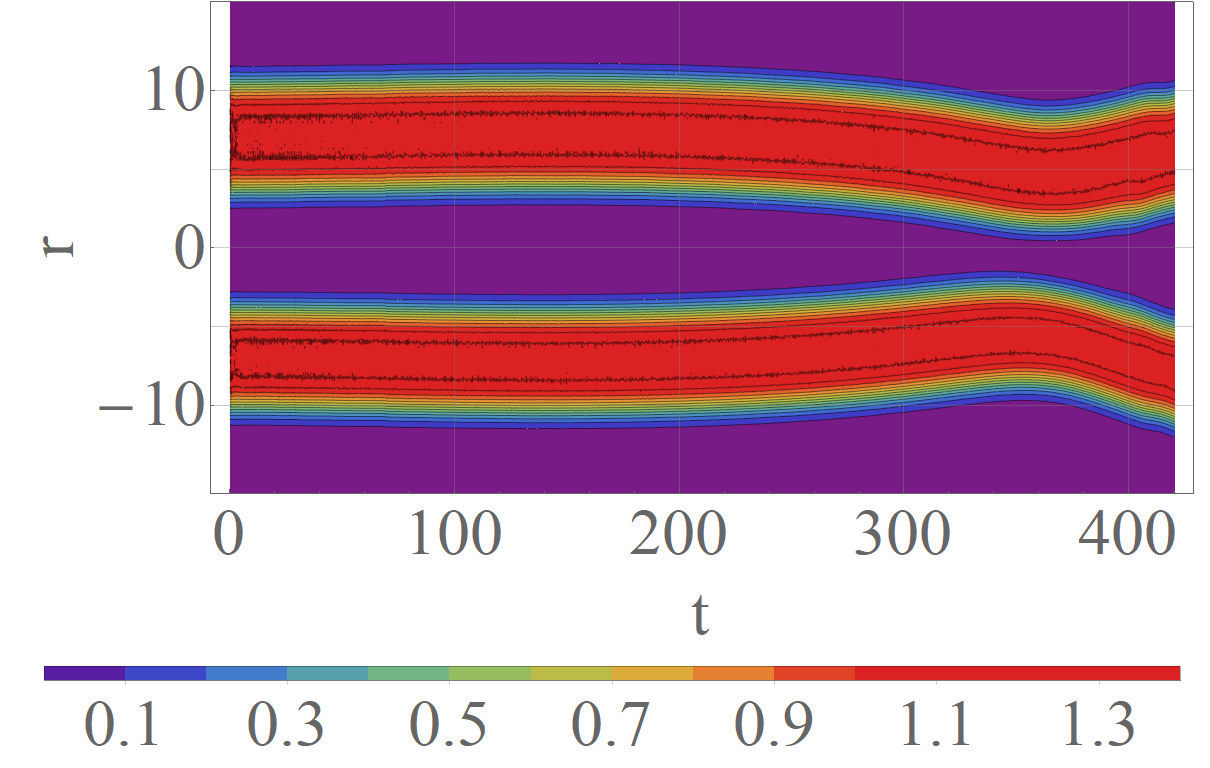}
     \caption{Comparison of collision of droplets with initially perturbed (right panel) and unperturbed (left panel) droplets.  Initial conditions: $N^L_1=203$, $N^R_1=162.4$, $\Delta \phi_1=\Delta \phi_2=3\pi/4$. Other parameters, dimension of axes and colour coding is the same as in Fig.1.}
      \label{phase_fluctuations}
  \end{figure}

The kinetic energy of the system after this perturbation is so large that the total energy of the droplets becomes positive, signifying thermal instability. Not only soft, but also hard modes are excited this way, and the phase difference between the two components varies from point to point. These phase fluctuations  convert into density fluctuations at early stages of the dynamics, on a time scale of about 0.05-0.1~ms. In Fig.(\ref{dens_diff}) relative density fluctuations of the right droplet (normalized to the non-perturbed density) are plotted. Fluctuations of the order of $\pm2\%$ are clearly visible. As densities depart locally from their equilibrium values, particles are emitted, which ultimately ceases both density and phase fluctuation. The phases stabilise at some values not significantly different from the initial setting (see right panel of Fig(\ref{dens_diff})). The internal dynamics of the droplets settle down, and the droplets continue their mutual approach until they eventually escape in opposite directions, see Fig(\ref{phase_fluctuations}).
The right panel shows dynamics of unperturbed droplets, and in the left panel a collision of perturbed droplets is illustrated. Both processes look very similar, however careful inspection suggests that the repulsion of perturbed droplets starts marginally later than in the analogous undisturbed setting. 
  
This example illustrates that phase fluctuations tend to self-stabilize. They are suppressed by emission of particles. The observed self-stabilization of large fluctuations of phases of droplets is a very interesting feature which deserves detailed study in the future.

\section{Conclusions}
\label{sec:7}
In conclusion, we showed that the dynamics of interacting droplets and their ultimate fate depend crucially on the relative phases of their wavefunctions. Two liquid quantum droplets, which constitute identical macroscopic objects, can be made to merge, repel or evaporate only by manipulating their quantum phases. Thus the processes studied in this work are macroscopic manifestations of the quantum nature of ultracold droplets. The interaction potential derived here as well as the two-component Josephson-junction equations may prove useful in studying the Andreev-Bashkin effect or modelling arrays of coupled droplets in a supersolid-like arrangement. Our Josephson-junction equations do not have any free parameters -- the chemical potentials and their derivatives as well as the long-range behaviour of the droplets' wavefunctions were found from stationary solutions of the extended Gross-Pitaevskii equations. The collisions discussed in this work enfold over a time span of more than 100 ms. The experimental verification of our predictions depends crucially on the realization of long-lived quantum droplets. One of the possible ways towards experiments are heteronuclear droplets in lower dimensions. On the other hand, one component Josephson junction dynamics is also possible in the case of dipolar droplets with sufficient lifetime. In fact Josephson-like oscillation of phases and atom number in droplets forming a supersolid were reported \cite{Pfau19}. Collisions of dipolar droplets remain to be studied.

We believe that experimental advances in creating long-living  droplets will sooner or later allow for verification of our predictions in their full extent.

\section{acknowledgments}
M. Pylak, F. Gampel and M. Gajda  acknowledge support from (Polish) National Science Centre grant No. 2017/25/B/ST2/01943. M. Gajda was also partially supported by: (Polish) National Science Center Grant No. DEC-2015/18/E/ST2/00760. F. Gampel acknowledges  project MAQS supported by the National Science Centre, Poland under QuantERA grant no. 2019/32/Z/ST2/00016, which has received funding from the European Union’s Horizon 2020 research and innovation program under grant agreement no 731473. M. P{\l}odzie\'n acknowledges the support of the Polish National Agency for Academic Exchange, the Bekker programme no: PPN/BEK/2020/1/00317. M. P{\l}odzie\'n   acknowledges also  support from Agencia Estatal de Investigación (the R\&D project CEX2019-000910-S, funded by MCIN/ AEI/10.13039/501100011033, Plan National FIDEUA PID2019-106901GB-I00, FPI), Fundació Privada Cellex, Fundació Mir-Puig, and from Generalitat de Catalunya (AGAUR Grant No. 2017 SGR 1341, CERCA program). 

\newpage
\bibliography{main}
\bibliographystyle{apsrev4-1}

\end{document}